\documentclass[superscriptaddress,reprint]{revtex4-1}

\usepackage{graphicx}
\usepackage{amssymb}
\usepackage{amsmath}
\usepackage{txfonts}
\usepackage{bm}
\usepackage{float}
\usepackage{natbib}
\setcitestyle{super}
\renewcommand{\bibnumfmt}[1]{#1.}

\makeatletter
\def\fnum@figure{\figurename\nobreakspace\textbf{\thefigure}}
\makeatother

\renewcommand{\figurename}{\textbf{Figure}}

\newcommand*\rfrac[2]{{}^{#1}\!/_{#2}}

\begin{document}

\small

\title{Polaritons in Living Systems: Modifying Energy Landscapes in Photosynthetic Organisms Using a Photonic Structure}

\author{David M Coles}
\email{d.m.coles@sheffield.ac.uk}
\affiliation{Department of Physics \& Astronomy, University of Sheffield, Sheffield S3 7RH, UK}

\author{Lucas C Flatten}
\affiliation{Department of Materials, University of Oxford, Oxford OX1 3PH, UK}

\author{Thomas Sydney}
\affiliation{Department of Chemistry, University of Sheffield, Sheffield S3 7HF, UK}

\author{Emily Hounslow}
\affiliation{Department of Chemical and Biological Engineering, University of Sheffield, Sheffield S1 3JD, UK}

\author{Semion K Saikin}
\affiliation{Department of Chemistry and Chemical Biology, Harvard University, Cambridge, MA 02138, USA}
\affiliation{Institute of Physics, Kazan Federal University Kazan, 420008, Russian Federation}

\author{Al\'{a}n Aspuru-Guzik}
\affiliation{Department of Chemistry and Chemical Biology, Harvard University, Cambridge, MA 02138, USA}

\author{Vlatko Vedral}
\affiliation{Department of Physics, University of Oxford, Oxford OX1 3PU, UK}
\affiliation{Centre for Quantum Technologies, National University of Singapore, Singapore 117543}

\author{Joseph Kuo-Hsiang Tang}
\altaffiliation{Currently at Materials and Manufacturing Directories, Air Force Research Laboratory, WPAFB, OH 45433, USA}
\affiliation{Department of Chemistry and Biochemistry, Clark University, Worcester, MA 01610-1477, USA}

\author{Robert A Taylor}
\affiliation{Department of Physics, University of Oxford, Oxford OX1 3PU, UK}

\author{Jason M Smith}
\email{jason.smith@materials.ox.ac.uk}
\affiliation{Department of Materials, University of Oxford, Oxford OX1 3PH, UK}

\author{David G Lidzey}
\email{d.g.lidzey@sheffield.ac.uk}
\affiliation{Department of Physics \& Astronomy, University of Sheffield, Sheffield S3 7RH, UK}

\maketitle

\textbf{
Photosynthetic organisms rely on a series of self-assembled nanostructures with tuned electronic energy levels in order to transport energy from where it is collected by photon absorption, to reaction centers where the energy is used to drive chemical reactions. In the photosynthetic bacteria \textit{Chlorobaculum tepidum} (\textit{Cba. tepidum}), a member of the green sulphur bacteria (GSB) family, light is absorbed by large antenna complexes called chlorosomes. The exciton generated is transferred to a protein baseplate attached to the chlorosome, before traveling through the Fenna-Matthews-Olson (FMO) complex to the reaction center\cite{overmann_the_prokaryotes}. The energy levels of these systems are generally defined by their chemical structure. Here we show that by placing bacteria within a photonic microcavity, we can access the strong exciton-photon coupling regime\cite{weisbuch_observation_of_the} between a confined cavity mode and exciton states of the chlorosome, whereby a coherent exchange of energy between the bacteria and cavity mode results in the formation of polariton states. The polaritons have an energy distinct from that of the exciton and photon, and can be tuned in situ via the microcavity length. This results in real-time, non-invasive control over the relative energy levels within the bacteria. This demonstrates the ability to strongly influence living biological systems with photonic structures such as microcavities. We believe that by creating polariton states, that are in this case a superposition of a photon and excitons within a living bacteria, we can modify energy transfer pathways and therefore study the importance of energy level alignment on the efficiency of photosynthetic systems.
}
\begin{figure}
\centering
\includegraphics[width=0.4\textwidth]{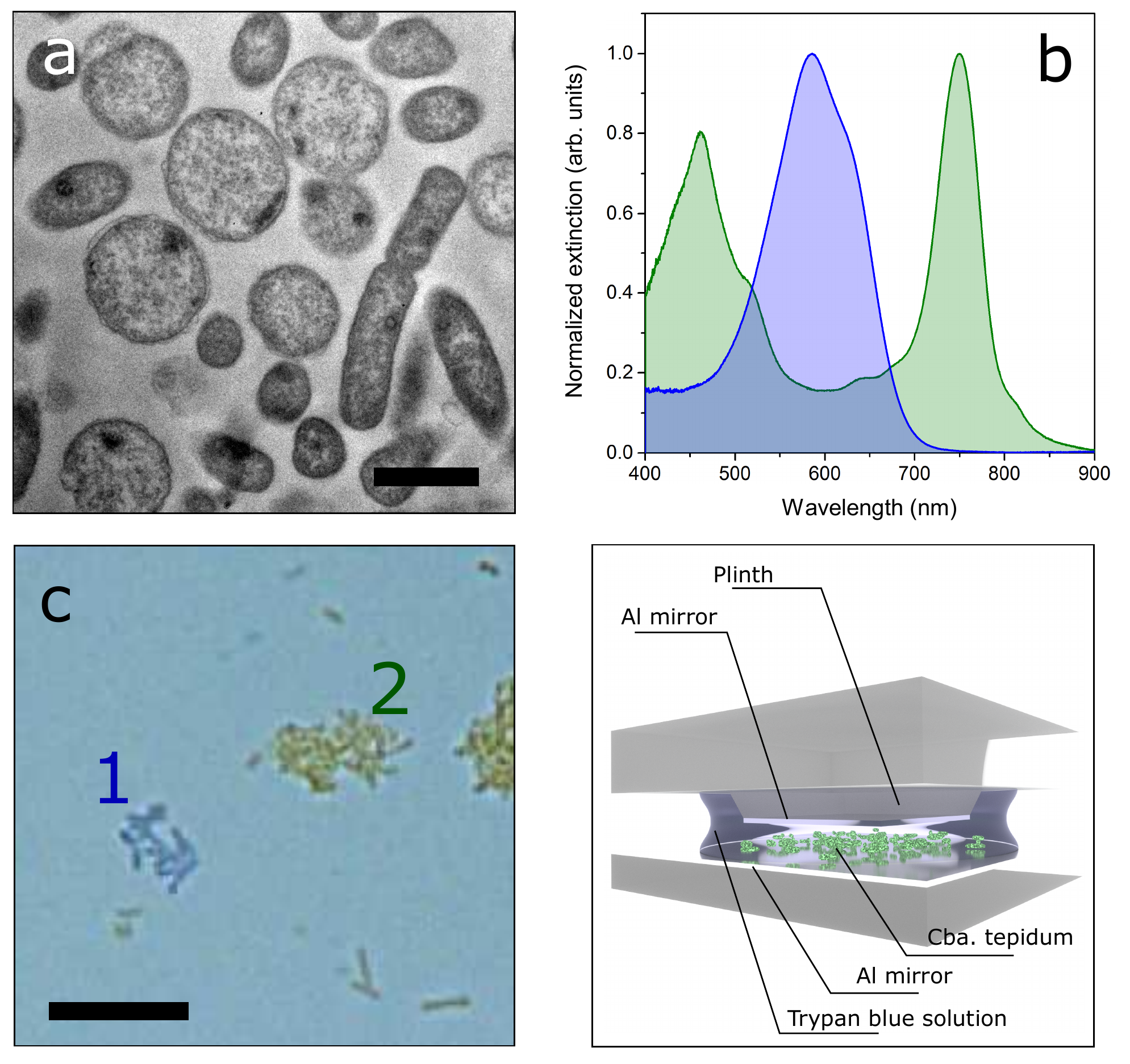}
\caption{\textbf{Spectral properties of green sulphur bacteria and microcavity configuration.}  \textbf{(a)} TEM image of \textit{Cba. tepidum}. Scale bar is 1 $\muup$m. Note that the size and shape of the bacteria is dependent upon the light conditions during growth\cite{saikin_chromic_acclimation_and}. \textbf{(b)} Normalised extinction spectra of 0.4\% trypan blue (TB) aqueous solution (blue line) and \textit{Cba. tepidum} in water (green line). \textbf{(c)} Optical microscope image of \textit{Cba. tepidum} in a TB viability stain showing clusters of bacteria with compromised cell membranes (stained blue, labeled \textbf{1}) and intact cell membranes (unstained, appear green, labeled \textbf{2}). The scale bar is 10 $\muup$m. \textbf{(d)} Schematic of microcavity consisting of a bacterial solution suspended between two semitransparent metallic mirrors, one of which is on a raised plinth.}
\label{fig1}
\end{figure}

A photonic structure has the ability to modify the properties of electronic transitions due to changes in the local density of photonic states. If both the resonator and exciton state display suitably low losses (i.e. narrow linewidths), the exciton has a strong interaction with light (i.e. a large absorption coefficient), and the resonator is degenerate in energy with the exciton, the system may enter the strong coupling regime. Here, energy is reversibly exchanged between the resonator and exciton and two new eigenstates of the system are formed which are a coherent superposition of the photonic and excitonic states. Such states are called polaritons, and are quasiparticles that are delocalized throughout the resonator due to their photonic component, whilst retaining an interaction cross-section inherited from the exciton\cite{exciton_polariton_condensates}. These properties have led to striking displays of phenomena such as polariton superfluids\cite{amo_collective_fluid_dynamics}, inversionless lasing\cite{schneider_an_electrically_pumped} and non-equilibrium Bose-Einstein condensates\cite{kasprzak_bose_einstein_condensation}, the latter two being observed up to room temperature (Refs. \citenum{bhattachary_room_temperature_electrically,plumhof_room_temperature_bose,daskalakis_nonlinear_interactions_in}).
Strong exciton-photon coupling in optical microcavities was first observed almost 25 years ago by Weisbuch et al. \cite{weisbuch_observation_of_the}, when a series of semiconductor quantum wells were embedded between two high quality planar dielectric mirrors. Since then, a wide variety of materials have been shown to be able to strongly couple to photonic modes such as bulk semiconductors\cite{vincent_observation_of_rabi}, organic molecules\cite{lidzey_strong_exciton_photon}, polymers\cite{takada_polariton_emission_from}, 2D dichalcogenides\cite{liu_strong_light_matter,flatten_room_temperature_exciton} and proteins\cite{dietrich_an_exciton_polariton}. Recently, single molecules were even shown to be able to strongly couple to plasmonic cavities\cite{chikkaraddy_single_molecule_strong}. In this Letter, we show that living bacteria placed within an optical microcavity are able to strongly couple to the cavity field. Our work therefore demonstrates the formation of a `living polariton'. This quasiparticle is part photon and part living organism in nature. This approach opens an opportunity to understand the interplay of the electronic states within a photosynthetic organism with its biological function. Looking further ahead, we expect that our approach will permit optical hybridisation of the chlorosome states with other optoelectronic materials, offering an entirely new way for the GSB to collect or deliver energy.

\onecolumngrid
\begin{center}
\begin{figure}[t]
\centering
\includegraphics[width=0.9\textwidth]{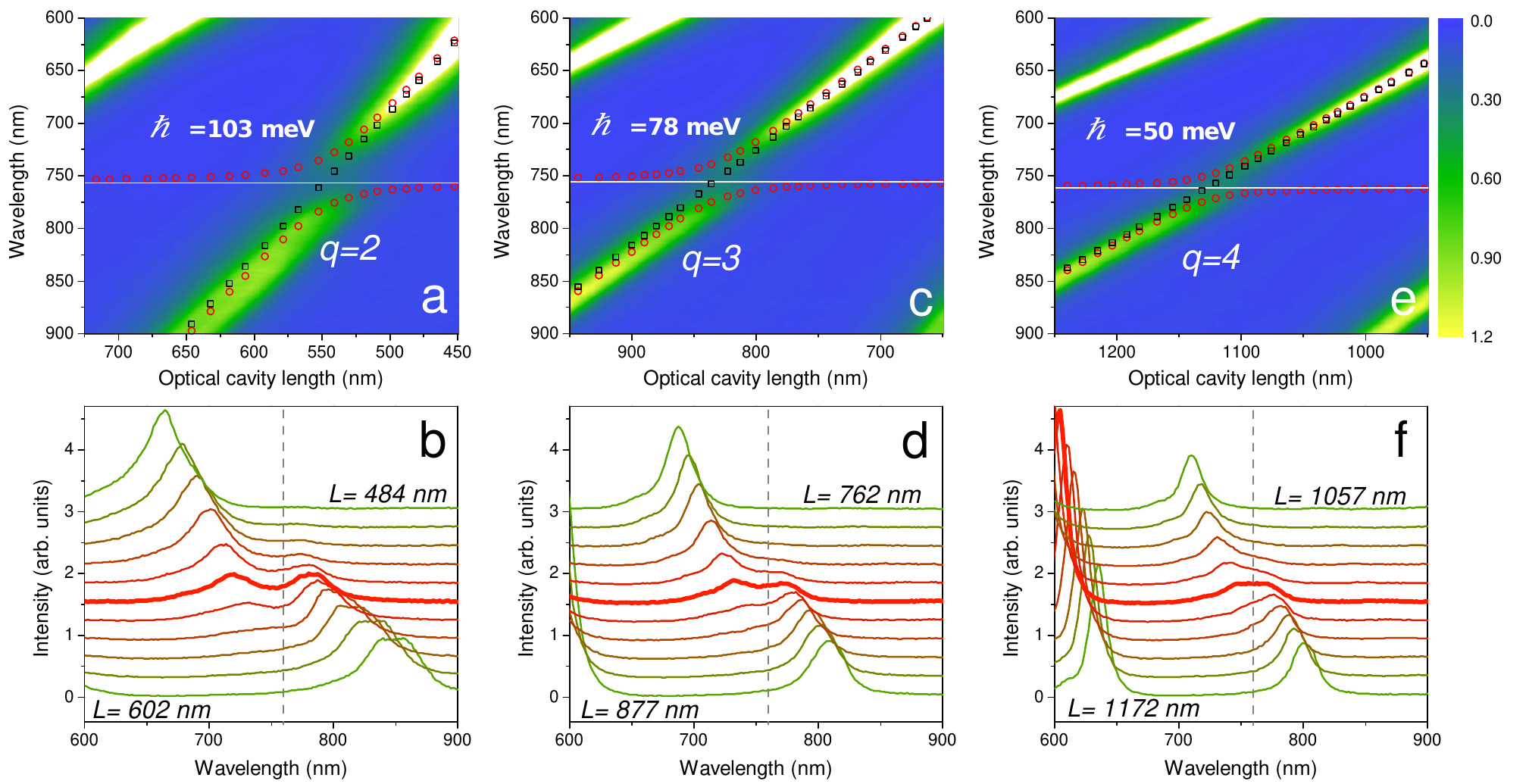}
\caption{\textbf{Strong coupling of green sulphur bacteria to microcavity photonic modes.}  \textbf{(a)} Transmission of the cavity at the point labeled \textbf{1} in Fig. \ref{fig3} as function of wavelength and cavity length while scanning the \textit{q=2} cavity mode through the chlorosome energy, showing the anticrossing of the polariton branches about the chlorosome energy. White horizontal line shows the chlorosome exciton energy and black squares show the unperturbed cavity mode energy. Red circles are the fitted polariton branch energies. \textbf{(b)} Individual transmission spectra, vertically offset, for given cavity lengths around exciton-photon resonance clearly showing the splitting of the cavity mode at exciton-photon resonance. Grey dashed line shows the chlorosome exciton energy. \textbf{(c)} and \textbf{(d)}, and \textbf{(e)} and \textbf{(f)} show the same for cavity modes $q=3$ and $q=4$ respectively.}
\label{fig2}
\end{figure}
\end{center}
\twocolumngrid

Figure \ref{fig1}(a) shows a TEM micrograph of a \textit{Cba. tepidum}. The bacteria were either grown following the procedure given in Ref. \citenum{overmann_the_prokaryotes}, or purchased as an active culture (Leibniz-Institut DSMZ), and were stored in anaerobic conditions prior to use. Each bacterial cell contains 200-250 light harvesting chlorosomes, which are large ovoid structures (100-200 nm long, $\sim$50 nm wide) consisting of tubular or planar aggregates of bacteriochlorophyll  \textit{c}  (BChl \textit{c})  molecules\cite{orf_chlorosome_antenna_complexes,oostergetel_the_chlorosome_a}. The absorption spectrum of \textit{Cba. tepidum} in aqueous solution (40 mg biomass per ml) is shown in Fig. \ref{fig1}(b) (green line). The strong absorption peak at 750 nm is due to aggregates of BChl \textit{c} in the chlorosomes. The weak absorption shoulder at 676 nm is assigned to BChl \textit{c} monomer and/or chlorophyll \textit{a} that can be found in the reaction center (while the principle exciton energy of the reaction center is at 840 nm). The shoulder at 810 nm is due to the FMO complex, while the absorption band in the 400-500 nm region is from the Soret band of BChl \textit{c} and carotenoid molecules within the chlorosomes. In order to verify the bacteria are alive in the strong coupling regime, we use the cell viability stain trypan blue (TB) which is added to the bacterial solution. The dye is able to permeate the cell membrane of dead cells and binds to intracellular proteins. Live cells with intact membranes are unstained by the dye\cite{herrera_validation_of_three}. The absorption spectrum of TB is shown in Fig. \ref{fig1}(b) (blue line), and displays a strong absorption peak at 587 nm, with a shoulder at 630 nm. Figure \ref{fig1}(c) shows an optical microscope image of the \textit{Cba. tepidum} solution stained with TB (0.4\% in water) at a ratio of 1:1. Both dead and alive clusters of bacteria are visible, labeled \textbf{1} and \textbf{2} respectively.

An open microcavity structure is used as the photonic resonator. Two 15 nm thick semitransparent aluminium planar mirrors (80\% reflectivity at 750 nm) were thermally deposited on silica substrates. One of the substrates has a raised `plinth' of dimensions 100 $\muup$m $\times$ 100 $\muup$m onto which the mirror is grown. A 10 nm layer of poly(methyl methacrylate) (PMMA) is spincast onto each mirror. The two mirrors are mounted face-to-face to form the cavity within a custom built white-light transmission microscope that allows angular alignment of the mirrors. A piezoelectric actuator allows nanometric control over the cavity length. The cavity is imaged onto the entrance slit of an imaging CCD spectrometer. The area to be spectrally imaged is defined by the image position on the spectrometer slit and the row of pixels on the CCD, the former defining the horizontal coordinate and the latter the vertical coordinate. The stained \textit{Cba. tepidum} solution is injected between the mirrors, before reducing the mirror separation to form a cavity with well-defined Fabry-Perot modes. A schematic of the cavity geometry is shown in figure \ref{fig1}(d). 

The transmission spectra from a region of the cavity measuring 5.5 $\muup$m $\times$ 1.2 $\muup$m as the cavity length is scanned from 450 nm to 725 nm is shown in figure \ref{fig2}(a) (see Supplementary Information for details on the calculation of the cavity length). The observed transmission peak corresponds to the $q=2$ cavity mode, where $q$ is the mode index. As the cavity mode energy is scanned through the exciton energy (solid white line), two peaks are observed that anticross about that energy. These peaks are the upper and lower polariton branches (UPB and LPB, red circles) that reside at higher and lower energy than the exciton respectively. While a strongly coupled system may be described using a fully quantum or semi-quantum formalism\cite{kavokin_microcavities}, here the large number of exciton states within the cavity allow us to use a classically coupled oscillator model\cite{cardona_light_scattering_in} to fit the polariton state energies (see Supplementary Information).

When the uncoupled photon and exciton energy are degenerate at the point of anticrossing, the polariton state can be considered 50\% photon and 50\% exciton. At this point, the magnitude of the energy splitting between the UPB and LPB is the Rabi splitting energy ($\hbar\Omega$) which is dependent on the square root of the product of the transition oscillator strength and number of states in the cavity mode volume. In the case of coupling with the $q=2$ cavity mode, we find a splitting of 103 meV. The criteria for strong coupling\cite{bajoni_polariton_lasers_hybrid} is that $\hbar\Omega > (\gamma_{\text{x}}/2)+(\gamma_{c}/2)$ where $\gamma_{\text{x}}$ and $\gamma_{c}$ are the full-width at half-maximum linewidths of the uncoupled exciton and photon respectively. The chlorosome exciton linewidth is 130 meV, and the $q=2$ cavity mode linewidth away from the strong coupling region is 70 meV, therefore the strong coupling criteria is satisfied for coupling to the $q=2$ mode.

Figure \ref{fig2}(b) shows a series of vertically offset transmission spectra for decreasing cavity length (bottom to top). The two polariton branches and their anticrossing about the exciton energy (grey dashed line) is clearly visible. We note that the cavity could not be closed beyond $\sim$ 450 nm, likely due to the size of the bacteria within the cavity.

Figure \ref{fig2}(c) and (d) show the microcavity transmission as the $q=3$ cavity mode is scanned through the exciton energy. While an anticrossing is again visible, the mode splitting is reduced to 78 meV. This is because of a reduction in the interaction potential due to the weaker EM-field within the cavity. For the $q=4$ mode (figure \ref{fig2}(e) and (f)), the splitting energy is reduced again to 50 meV, and the anticrossing is not clearly resolvable.

The magnitude of the Rabi splitting allows us to put bounds on the number of pigment molecules, and hence the number of bacteria involved in the coupling (see Supplementary Information for details). We find that the number of excitons simultaneously coupled to the $q=2$ cavity mode is $\sim$95 million if all chlorosomes are oriented in the plane of the cavity, and $\sim$220 million if all dipoles are randomly oriented in the cavity. Assuming 200,000 BChl molecules per chlorosome, the splitting corresponds to the coupling of excitons from between 470 and 1100 chlorosomes, approximately the number that are in 2 to 6 bacteria.

\begin{figure}
\centering
\includegraphics[width=0.45\textwidth]{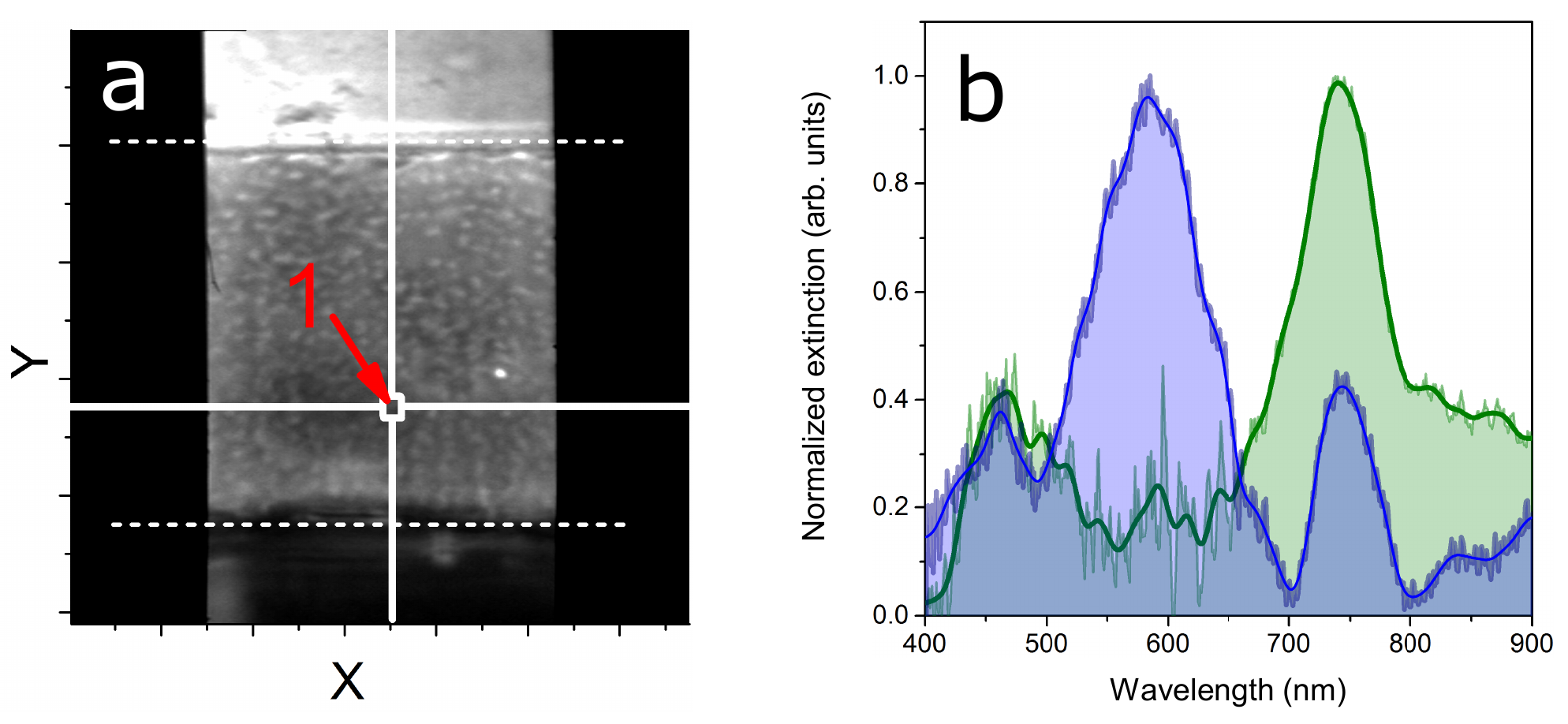}
\caption{\textbf{\textit{Cba. tepidum}  within a microcavity and microabsorption of \textit{Cba. tepidum}.}  (\textbf{a}) Real space optical image of the microcavity. White dashed lines mark the extent of the plinth in the vertical direction. White solid vertical line represents the position of the spectrometer slit when performing spectral imaging. White solid horizontal line represents the position of the CCD track used for the spectra showing strong coupling shown in figure \ref{fig2}. The intersection of the solid lines (marked \textbf{1}) is the position of the bacteria that are shown to undergo strong coupling to the cavity. The bacteria appear as pale spots on the image, however they are not clearly individually resolvable as they are smaller than the resolution of the microscope. (\textbf{b}) Normalized absorption spectrum taken at position \textbf{1} when the cavity was opened to allow a continuum of photonic states (green line), and normalized absorption spectrum of stained bacteria (blue line). In both cases, the absorption spectra were taken from a spectral image, with the reference taken from the same image but a separate track where no bacteria are present.}
\label{fig3}
\end{figure}

In order to ascertain whether the bacteria are alive during strong coupling, we have performed micro-extinction spectroscopy on the bacteria involved in the coupling. A real-space CCD image of the cavity is shown in figure \ref{fig3}(a). The cavity was opened to approximately 100 $\muup$m to allow a continuum of photonic states, and the normalized extinction spectrum of the region marked `\textbf{1}' in Fig. \ref{fig3}(a) is shown in figure \ref{fig3}(b) (green line). We see that there is a strong absorption peak at 750 nm due to the chlorosome absorption, but no sign of TB absorption in the 500-650 nm range, indicating that the cells had not been stained and remained viable. For comparison, the micro-extinction spectrum of an area containing compromised bacteria is also shown (blue line) where TB absorption is the dominant feature. Furthermore, there is no apparent dip in transmission intensity of the cavity modes when scanning through the 500-600 nm range that would indicate the presence of TB (Supplementary Fig. S1). While the cavity acts to restrict the intensity of light reaching the bacteria, they are known to survive in extremely low light environments and display a low mortality rate even in the presence of no light\cite{blankenship_anoxygenic_photosynthetic_bacteria}, making long-term experiments based on bacterial growth rates feasible. Indeed, the bacteria under investigation remained unstained for the duration of the experiment, totaling several hours.

We have previously suggested that the polariton branches may provide an alternative pathway for excitons to migrate through the photosynthetic system, bypassing various states\cite{coles_strong_coupling_between}. The baseplate energy is at 790 nm, while the FMO and reaction center are positioned at 810 nm and 840 nm respectively. Here, the lower polariton branch energy can be widely tuned via the cavity length, and can be brought into resonance with each of these structures. For coupling to the $q=2$ cavity mode, the LPB is resonant with the baseplate at a cavity length of $L= $560 nm, the FMO complex at $L= $580 nm and the reaction center at $L= $605 nm. The excitonic percentage of polariton states at the BP, FMO and reaction center energy are 32\%, 18\% and 8\% respectively. The LPB may therefore act as a relaxation pathway for excitons from the chlorosome directly into lower energy states, including directly to the reaction center. This should modify the energy transfer rates between the chlorosome and the other subunits \cite{huh_atomistic_study_of} and as a consequence affect the growth rate of the bacteria.

In conclusion, we have introduced living photosynthetic bacteria into a photonic microcavity and shown that the system can enter the strong coupling regime, thus creating exciton-photon superposition states within a living organism.  It opens the opportunity to create hybrid-polariton systems in which the optical state that is coupled to the chlorosome assembly is also coupled to a second semiconductor material placed within the optical cavity. This approach has previously been used to hybridise a range of different semiconductor systems, including different species of molecular dyes\cite{lidzey_photon_mediated_hybridization}, and molecular dyes with semiconductor quantum wells\cite{holmes_strong_coupling_and,wenus_hybrid_organic_inorganic}. Such hybridisation has been shown to facilitate rapid energy transfer between the excitonic states by virtue of the intermediate hybrid-polariton\cite{coles_polariton_mediated_energy}, and could be used  to either inject or extract energy from a chlorosome in a living bacteria. Furthermore, the optical cavity allows in situ control over the relative energy levels within the bacteria, and by enhancing energy transfer to the reaction center from the chlorosome, it may be possible to direct the evolution of green-sulfur-bacteria towards organisms that are more fit to live inside a microcavity than outside of it, i.e. an organism tailored to live in a superposition state with a photon.


\begin{thebibliography}{10}
\expandafter\ifx\csname url\endcsname\relax
  \def\url#1{\texttt{#1}}\fi
\expandafter\ifx\csname urlprefix\endcsname\relax\def\urlprefix{URL }\fi
\providecommand{\bibinfo}[2]{#2}
\providecommand{\eprint}[2][]{\url{#2}}

\bibitem{overmann_the_prokaryotes}
\bibinfo{author}{Overmann, J.}
\newblock \emph{\bibinfo{title}{The Prokaryotes}}, vol.~\bibinfo{volume}{7}
  (\bibinfo{publisher}{Springer, New York}, \bibinfo{year}{2006}),
  \bibinfo{edition}{3rd} edn.

\bibitem{weisbuch_observation_of_the}
\bibinfo{author}{Weisbuch, C.}, \bibinfo{author}{Nishioka, M.},
  \bibinfo{author}{Ishikawa, A.} \& \bibinfo{author}{Arakawa, Y.}
\newblock \bibinfo{title}{Observation of the coupled exciton-photon mode
  splitting in a semiconductor quantum microcavity}.
\newblock \emph{\bibinfo{journal}{Phys. Rev. Lett.}}
  \textbf{\bibinfo{volume}{69}}, \bibinfo{pages}{3314--3317}
  (\bibinfo{year}{1992}).

\bibitem{saikin_chromic_acclimation_and}
\bibinfo{author}{Saikin, S.~K.} \emph{et~al.}
\newblock \bibinfo{title}{Chromatic acclimation and population dynamics of
  green sulfur bacteria grown with spectrally tailored light}.
\newblock \emph{\bibinfo{journal}{Scientific Reports}}
  \textbf{\bibinfo{volume}{4}}, \bibinfo{pages}{5057} (\bibinfo{year}{2014}).

\bibitem{exciton_polariton_condensates}
\bibinfo{author}{Byrnes, T.}, \bibinfo{author}{Kim, N.~Y.} \&
  \bibinfo{author}{Yamamoto, Y.}
\newblock \bibinfo{title}{Exciton–polariton condensates}.
\newblock \emph{\bibinfo{journal}{Nature Physics}}
  \textbf{\bibinfo{volume}{10}}, \bibinfo{pages}{803--813}
  (\bibinfo{year}{2014}).

\bibitem{amo_collective_fluid_dynamics}
\bibinfo{author}{Amo, A.} \emph{et~al.}
\newblock \bibinfo{title}{Collective fluid dynamics of a polariton condensate
  in a semiconductor microcavity}.
\newblock \emph{\bibinfo{journal}{Nature}} \textbf{\bibinfo{volume}{457}},
  \bibinfo{pages}{291} (\bibinfo{year}{2009}).

\bibitem{schneider_an_electrically_pumped}
\bibinfo{author}{Schneider, C.} \emph{et~al.}
\newblock \bibinfo{title}{An electrically pumped polariton laser}.
\newblock \emph{\bibinfo{journal}{Nature}} \textbf{\bibinfo{volume}{497}},
  \bibinfo{pages}{348--352} (\bibinfo{year}{2013}).

\bibitem{kasprzak_bose_einstein_condensation}
\bibinfo{author}{Kasprzak, J.} \emph{et~al.}
\newblock \bibinfo{title}{Bose-einstein condensation of exciton polaritons}.
\newblock \emph{\bibinfo{journal}{Nature}} \textbf{\bibinfo{volume}{443}},
  \bibinfo{pages}{409} (\bibinfo{year}{2006}).

\bibitem{bhattachary_room_temperature_electrically}
\bibinfo{author}{Bhattacharya, P.} \emph{et~al.}
\newblock \bibinfo{title}{Room temperature electrically injected polariton
  laser}.
\newblock \emph{\bibinfo{journal}{Phys. Rev. Lett.}}
  \textbf{\bibinfo{volume}{112}}, \bibinfo{pages}{236802}
  (\bibinfo{year}{2014}).

\bibitem{plumhof_room_temperature_bose}
\bibinfo{author}{Plumhof, J.~D.}, \bibinfo{author}{St\"{o}ferle, T.},
  \bibinfo{author}{Mai, L.}, \bibinfo{author}{Scherf, U.} \&
  \bibinfo{author}{Mahrt, R.~F.}
\newblock \bibinfo{title}{Room-temperature {B}ose-{E}instein condensation of
  cavity exciton-polaritons in a polymer}.
\newblock \emph{\bibinfo{journal}{Nature Materials}}
  \textbf{\bibinfo{volume}{13}}, \bibinfo{pages}{247--252}
  (\bibinfo{year}{2014}).

\bibitem{daskalakis_nonlinear_interactions_in}
\bibinfo{author}{Daskalakis, K.~S.}, \bibinfo{author}{Maier, S.~A.},
  \bibinfo{author}{Murray, R.} \& \bibinfo{author}{K\'{e}na-Cohen, S.}
\newblock \bibinfo{title}{Nonlinear interactions in an organic polariton
  condensate}.
\newblock \emph{\bibinfo{journal}{Nature Materials}}
  \textbf{\bibinfo{volume}{13}}, \bibinfo{pages}{271--278}
  (\bibinfo{year}{2014}).

\bibitem{vincent_observation_of_rabi}
\bibinfo{author}{Antoine-Vincent, N.} \emph{et~al.}
\newblock \bibinfo{title}{Observation of rabi splitting in a bulk gan
  microcavity grown on silicon}.
\newblock \emph{\bibinfo{journal}{Phys. Rev. B}} \textbf{\bibinfo{volume}{68}},
  \bibinfo{pages}{153313} (\bibinfo{year}{2003}).

\bibitem{lidzey_strong_exciton_photon}
\bibinfo{author}{Lidzey, D.~G.} \emph{et~al.}
\newblock \bibinfo{title}{Strong exciton-photon coupling in an organic
  semiconductor microcavity}.
\newblock \emph{\bibinfo{journal}{Nature}} \textbf{\bibinfo{volume}{395}},
  \bibinfo{pages}{53--55} (\bibinfo{year}{1998}).

\bibitem{takada_polariton_emission_from}
\bibinfo{author}{Takada, N.}, \bibinfo{author}{Kamata, T.} \&
  \bibinfo{author}{Bradley, D. D.~C.}
\newblock \bibinfo{title}{Polariton emission from polysilane-based organic
  microcavities}.
\newblock \emph{\bibinfo{journal}{Applied Physics Letters}}
  \textbf{\bibinfo{volume}{82}}, \bibinfo{pages}{1812--1814}
  (\bibinfo{year}{2003}).

\bibitem{liu_strong_light_matter}
\bibinfo{author}{Liu, X.} \emph{et~al.}
\newblock \bibinfo{title}{Strong light–matter coupling in two-dimensional
  atomic crystals}.
\newblock \emph{\bibinfo{journal}{Nature Photonics}}
  \textbf{\bibinfo{volume}{9}}, \bibinfo{pages}{30--34} (\bibinfo{year}{2015}).

\bibitem{flatten_room_temperature_exciton}
\bibinfo{author}{Flatten, L.~C.} \emph{et~al.}
\newblock \bibinfo{title}{Room-temperature exciton-polaritons with
  two-dimensional ws2} \bibinfo{pages}{arXiv:1605.04743v2}
  (\bibinfo{year}{2016}).

\bibitem{dietrich_an_exciton_polariton}
\bibinfo{author}{Dietrich, C.~P.} \emph{et~al.}
\newblock \bibinfo{title}{An exciton-polariton laser based on biologically
  produced fluorescent protein}.
\newblock \emph{\bibinfo{journal}{Science Advances}}
  \textbf{\bibinfo{volume}{2}}, \bibinfo{pages}{1--7} (\bibinfo{year}{2016}).

\bibitem{chikkaraddy_single_molecule_strong}
\bibinfo{author}{Chikkaraddy, R.} \emph{et~al.}
\newblock \bibinfo{title}{Single-molecule strong coupling at room temperature
  in plasmonic nanocavities}.
\newblock \emph{\bibinfo{journal}{Nature}} \textbf{\bibinfo{volume}{535}},
  \bibinfo{pages}{127–130} (\bibinfo{year}{2016}).

\bibitem{orf_chlorosome_antenna_complexes}
\bibinfo{author}{Orf, G.~S.} \& \bibinfo{author}{Blankenship, R.~E.}
\newblock \bibinfo{title}{Chlorosome antenna complexes from green
  photosynthetic bacteria}.
\newblock \emph{\bibinfo{journal}{Photosynthesis Research}}
  \textbf{\bibinfo{volume}{116}}, \bibinfo{pages}{315--331}
  (\bibinfo{year}{2013}).

\bibitem{oostergetel_the_chlorosome_a}
\bibinfo{author}{Oostergetel, G.}, \bibinfo{author}{Amerongen, H.} \&
  \bibinfo{author}{Boekema, E.}
\newblock \bibinfo{title}{The chlorosome: a prototype for efficient light
  harvesting in photosynthesis}.
\newblock \emph{\bibinfo{journal}{Photosynthesis Research}}
  \textbf{\bibinfo{volume}{104}}, \bibinfo{pages}{245--255}
  (\bibinfo{year}{2010}).

\bibitem{herrera_validation_of_three}
\bibinfo{author}{Cadena-Herrera, D.} \emph{et~al.}
\newblock \bibinfo{title}{Validation of three viable-cell counting methods:
  Manual, semi-automated, and automated}.
\newblock \emph{\bibinfo{journal}{Biotechnology Reports}}
  \textbf{\bibinfo{volume}{7}}, \bibinfo{pages}{9 -- 16}
  (\bibinfo{year}{2015}).

\bibitem{kavokin_microcavities}
\bibinfo{author}{Kavokin, A.~V.}, \bibinfo{author}{Baumberg, J.},
  \bibinfo{author}{Malpuech, G.} \& \bibinfo{author}{Laussy, F.}
\newblock \emph{\bibinfo{title}{Microcavities}} (\bibinfo{publisher}{Oxford
  University Press}, \bibinfo{year}{2007}).

\bibitem{cardona_light_scattering_in}
\bibinfo{author}{Cardona, M.} \& \bibinfo{author}{Merlin, R.}
\newblock \emph{\bibinfo{title}{Light Scattering in Solids IX: Novel Materials
  and Techniques}} (\bibinfo{publisher}{Springer}, \bibinfo{year}{2007}).

\bibitem{bajoni_polariton_lasers_hybrid}
\bibinfo{author}{Bajoni, D.}
\newblock \bibinfo{title}{Polariton lasers. hybrid light–matter lasers
  without inversion}.
\newblock \emph{\bibinfo{journal}{Journal of Physics D: Applied Physics}}
  \textbf{\bibinfo{volume}{45}}, \bibinfo{pages}{313001}
  (\bibinfo{year}{2012}).

\bibitem{blankenship_anoxygenic_photosynthetic_bacteria}
\bibinfo{author}{Blankenship, R.}, \bibinfo{author}{Madigan, M.} \&
  \bibinfo{author}{Bauer, C.}
\newblock \emph{\bibinfo{title}{Anoxygenic Photosynthetic Bacteria}}
  (\bibinfo{publisher}{Springer}, \bibinfo{year}{1995}).

\bibitem{coles_strong_coupling_between}
\bibinfo{author}{Coles, D.~M.} \emph{et~al.}
\newblock \bibinfo{title}{Strong coupling between chlorosomes of photosynthetic
  bacteria and a conﬁned optical cavity mode}.
\newblock \emph{\bibinfo{journal}{Nature Communications}}
  \textbf{\bibinfo{volume}{5}}, \bibinfo{pages}{5561} (\bibinfo{year}{2014}).

\bibitem{huh_atomistic_study_of}
\bibinfo{author}{Huh, J.} \emph{et~al.}
\newblock \bibinfo{title}{Atomistic study of energy funneling in the
  light-harvesting complex of green sulfur bacteria}.
\newblock \emph{\bibinfo{journal}{Journal of the American Chemical Society}}
  \textbf{\bibinfo{volume}{136}}, \bibinfo{pages}{2048--2057}
  (\bibinfo{year}{2014}).

\bibitem{lidzey_photon_mediated_hybridization}
\bibinfo{author}{Lidzey, D.~G.}, \bibinfo{author}{Bradley, D. D.~C.},
  \bibinfo{author}{Armitage, A.}, \bibinfo{author}{Walker, S.} \&
  \bibinfo{author}{Skolnick, M.~S.}
\newblock \bibinfo{title}{Photon-mediated hybridization of frenkel excitons in
  organic semiconductor microcavities}.
\newblock \emph{\bibinfo{journal}{Science}} \textbf{\bibinfo{volume}{288}},
  \bibinfo{pages}{1620--1623} (\bibinfo{year}{2000}).

\bibitem{holmes_strong_coupling_and}
\bibinfo{author}{Holmes, R.~J.}, \bibinfo{author}{K\'ena-Cohen, S.},
  \bibinfo{author}{Menon, V.~M.} \& \bibinfo{author}{Forrest, S.~R.}
\newblock \bibinfo{title}{Strong coupling and hybridization of frenkel and
  wannier-mott excitons in an organic-inorganic optical microcavity}.
\newblock \emph{\bibinfo{journal}{Phys. Rev. B}} \textbf{\bibinfo{volume}{74}},
  \bibinfo{pages}{235211} (\bibinfo{year}{2006}).

\bibitem{wenus_hybrid_organic_inorganic}
\bibinfo{author}{Wenus, J.} \emph{et~al.}
\newblock \bibinfo{title}{Hybrid organic-inorganic exciton-polaritons in a
  strongly coupled microcavity}.
\newblock \emph{\bibinfo{journal}{Phys. Rev. B}} \textbf{\bibinfo{volume}{74}},
  \bibinfo{pages}{235212} (\bibinfo{year}{2006}).

\bibitem{coles_polariton_mediated_energy}
\bibinfo{author}{Coles, D.~M.} \emph{et~al.}
\newblock \bibinfo{title}{Polariton-mediated energy transfer between organic
  dyes in a strongly coupled optical microcavity}.
\newblock \emph{\bibinfo{journal}{Nature Materials}}
  \textbf{\bibinfo{volume}{13}}, \bibinfo{pages}{712--719}
  (\bibinfo{year}{2014}).

\end{thebibliography}

\begin{thebibliography}{1}
\expandafter\ifx\csname url\endcsname\relax
  \def\url#1{\texttt{#1}}\fi
\expandafter\ifx\csname urlprefix\endcsname\relax\def\urlprefix{URL }\fi
\providecommand{\bibinfo}[2]{#2}
\providecommand{\eprint}[2][]{\url{#2}}

\bibitem{skolnick_strong_coupling_phenomena}
\bibinfo{author}{Skolnick, M.~S.}, \bibinfo{author}{Fisher, T.~A.} \&
  \bibinfo{author}{Whittaker, D.~M.}
\newblock \bibinfo{title}{Strong coupling phenomena in quantum microcavity
  structures}.
\newblock \emph{\bibinfo{journal}{Semicond. Sci. Technol.}}
  \textbf{\bibinfo{volume}{13}}, \bibinfo{pages}{645} (\bibinfo{year}{1998}).

\bibitem{liu_real_time_measurement}
\bibinfo{author}{Liu, P.} \emph{et~al.}
\newblock \bibinfo{title}{Real-time measurement of single bacterium's
  refractive index using optofluidic immersion refractometry}.
\newblock \emph{\bibinfo{journal}{Procedia Engineering}}
  \textbf{\bibinfo{volume}{87}}, \bibinfo{pages}{356 -- 359}
  (\bibinfo{year}{2014}).

\bibitem{fox_quantum_optics_an}
\bibinfo{author}{Fox, M.}
\newblock \emph{\bibinfo{title}{Quantum Optics - An Introduction}}
  (\bibinfo{publisher}{Oxford Master Series}, \bibinfo{year}{2006}).

\bibitem{prokhorenko_exciton_dynamics_in}
\bibinfo{author}{Prokhorenko, V.}, \bibinfo{author}{Steensgaard, D.} \&
  \bibinfo{author}{Holzwarth, A.}
\newblock \bibinfo{title}{Exciton dynamics in the chlorosomal antennae of the
  green bacteria chloroflexus aurantiacus and chlorobium tepidum}.
\newblock \emph{\bibinfo{journal}{Biophysical Journal}}
  \textbf{\bibinfo{volume}{79}}, \bibinfo{pages}{2105 -- 2120}
  (\bibinfo{year}{2000}).

\bibitem{ujihara_spontaneous_emission_and}
\bibinfo{author}{Ujihara, K.}
\newblock \bibinfo{title}{Spontaneous emission and the concept of effective
  area in a very short optical cavity with plane-parallel dielectric mirrors}.
\newblock \emph{\bibinfo{journal}{Japanese Journal of Applied Physics}}
  \textbf{\bibinfo{volume}{30}}, \bibinfo{pages}{L901--L903}
  (\bibinfo{year}{1991}).

\end{thebibliography}

{\footnotesize
\subsection*{Acknowledgements}
D.M.C and D.G.L thank EPSRC grant number EP/M025330/1 "Hybrid Polaritonics". D.M.C, R.A.T and J.M.S gratefully acknowledge the Oxford Martin School. L.C.F acknowledges funding from the Leverhulme Trust. S.K.S acknowledges the support from the subsidy allocated to Kazan Federal University for performing the state assignment in the area of scientific activities. A.A.G thanks the Center for Excitonics, an Energy Frontier Research Center funded by the U.S. Department of Energy, Office of Science and Office of Basic Energy Sciences, under Award Number DE-SC0001088. V.V thanks the Oxford Martin School, Wolfson College and the University of Oxford, the Leverhulme Trust (UK), the John Templeton Foundation, the EU Collaborative Project TherMiQ (Grant Agreement 618074), the COST Action MP1209 and the EPSRC (UK). This research is also supported by the National Research Foundation, Prime Minister’s Office, Singapore, under its Competitive Research Programme (CRP Award No. NRF- CRP14-2014-02) and administered by Centre for Quantum Technologies, National University of Singapore. D.M.C further thanks EPSRC grant number EP/K032518/1 "Characterisation of Nanomaterials for Energy".
}

{\footnotesize
\subsection*{Author Contributions}

S.K.S, A.A.G and V.V conceived the experiment and provided theoretical background. J.K.T cultured bacteria. Bacterial staining and optical microscopy was performed by T.S and E.H. D.M.C and L.C.F fabricated the microcavity samples. Experiments were performed by D.M.C and L.C.F under the supervision of R.A.T, J.M.S and D.G.L. All authors contributed to the preparation of the manuscript.
}



\renewcommand{\theequation}{S\arabic{equation}}
\renewcommand{\thefigure}{S\arabic{figure}}
\renewcommand*{\citenumfont}[1]{S#1}
\renewcommand*{\bibnumfmt}[1]{[S#1]}

\clearpage

\section*{Supplementary Information}

\subsection*{Calibration of cavity length}

The cavity mirrors are first aligned to be parallel by observing the interference fringes in transmission as the cavity length is reduced, and adjusting the angle of one of the mirrors until only one fringe is visible across the plinth. The cavity length is scanned by applying a linear voltage ramp to a piezoelectric actuator attached to one of the cavity mirrors, and the cavity transmission was spectrally imaged as a function of time. In order to calculate the cavity length for each spectrum, an area of the spectral image is selected where no strong coupling is observed (i.e. there is no bacteria or too few in a given area to couple). The cavity length ($L$) is set such that several Fabry-Perot modes are observed in transmission. The index ($q$) of each of the modes is calculated from the wavelengths of adjacent modes $\lambda_{q}$ and $\lambda_{q-1}$ using
\begin{equation}
q=\frac{\lambda_{q-1}}{\lambda_{q-1}-\lambda_{q}}
\label{mode_num}
\end{equation}

Once the mode index of a transmission peak is known, the cavity length can be calculated via 

\begin{equation}
L=\frac{\lambda_{q}q}{2n}
\label{cav_length}
\end{equation}
where $n$ is the intracavity refractive index, which in this case is that of water, 1.33.

\subsection*{Polariton branch fitting}

The energy of the lower and upper polariton branches is given by a coupled oscillator model\cite{skolnick_strong_coupling_phenomena}, equation \ref{coupled_oscillator}.

\begin{equation}
\left( \begin{array}{cc}
E_{c}(L) & \rfrac{\hbar\Omega}{2} \\
\rfrac{\hbar\Omega}{2} & E_{\text{x}} \end{array} \right) \left( \begin{array}{c}
\alpha_{c}(L) \\
\alpha_{\text{x}}(L) \end{array} \right) = E_{\text{p}}(L) \left( \begin{array}{c}
\alpha_{c}(L) \\
\alpha_{\text{x}}(L) \end{array} \right),
\label{coupled_oscillator}
\end{equation}
where $E_{c}(L)$ is the uncoupled cavity mode energy (which is found from Eqn. \ref{cav_length}), $E_{\text{x}} $ is the exciton energy, $E_{\text{p}}(L)$ is the polariton energy and $\alpha_{c}(L)$ and $\alpha_{\text{x}}(L)$ are photon and exciton mixing coefficients respectively. This can be solved to give

\begin{equation}
E_{p}(L)=\frac{E_{c}(L)+E_{\text{x}}}{2} \pm \sqrt{\left( E_{c}(L) - E_{\text{x}} \right)^{2}+\hbar\Omega^{2}}
\label{branch_energies}
\end{equation}
This is fitted to the observed polariton branch energies with $\hbar\Omega$ and $n$ as a fitting parameters. We note that refractive index of the bacteria varies slightly from that of the background index of water\cite{liu_real_time_measurement}, and from the fitting we find that $n=1.38$, $n=1.36$ and $n=1.35$ for the $q=2$, $q=3$ and $q=4$ cavity modes respectively, commensurate with a larger fraction of the cavity length being occupied by bacteria for smaller cavity lengths leading to a slightly higher value for $n$. The eigenvalues (mixing co-efficients) $\alpha_{c}(L)$ and $\alpha_{\text{x}}(L)$ can also be calculated via

\begin{align}
\alpha_{c,UPB}^{2}(L)=\frac{E_{c}(L)-E_{\text{p,UPB}}(L)}{E_{\text{x}}+E_{c}(L)-2E_{\text{p,UPB}}(L)}, \hspace{0.5cm} \alpha_{\text{x,UPB}}^{2}(L)=1-\alpha_{c,\text{UPB}}^{2}(L) \nonumber \\
\alpha_{c,LPB}^{2}(L)=\alpha_{\text{x,UPB}}^{2}(L), \hspace{1cm} \alpha_{\text{x,LPB}}^{2}(L)=\alpha_{c,UPB}^{2}(L)
\label{mixing_coefficients}
\end{align}

\subsection*{Transmission mode intensity}

\begin{center}
\begin{figure}[h]
\centering
\includegraphics[width=0.45\textwidth]{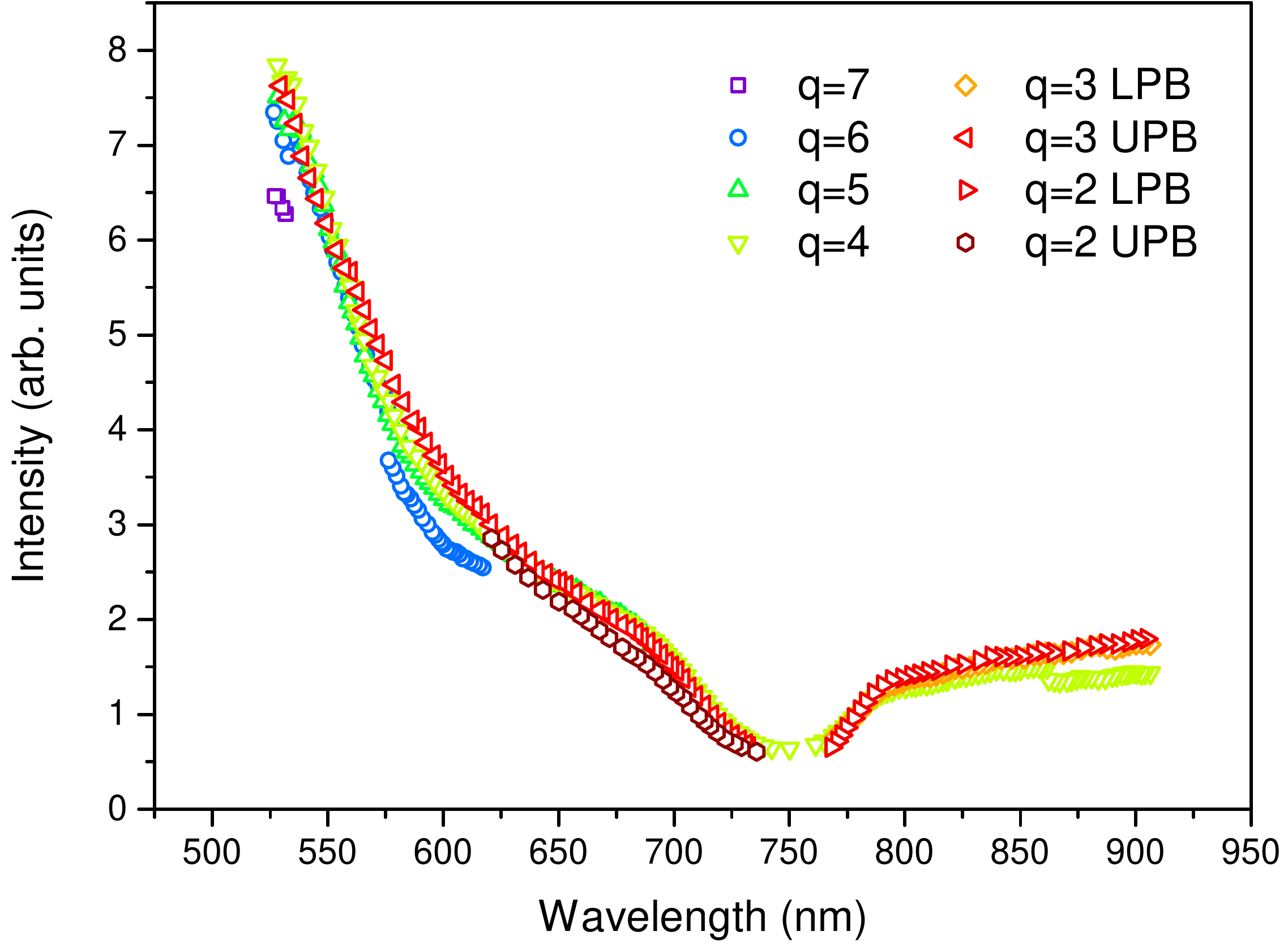}
\caption{\textbf{Mode intensity of polariton branches.} }
\label{figs1}
\end{figure}
\end{center}

Figure \ref{figs1} shows the transmission intensity for modes $q=2$ to $q=6$. We observe no obvious absorption feature at 587 nm that would correspond to the TB absorption band. The continuous fall in transmission intensity for increasing wavelength towards the exciton energy is due to the polariton branch becoming more exciton like and less photon like in nature, while the opposite is true for increasing wavelengths away from the exciton energy.

\subsection*{Calculation of the number of dipoles involved in strong coupling}

The Rabi splitting energy ($\hbar\Omega$) increases as the square root of the number of dipoles coupled to the field ($N$), with the splitting given by\cite{fox_quantum_optics_an}

\begin{equation}
\hbar\Omega=2\left(\bm{\muup}\cdot\hat{\textbf{E}}\right)\sqrt{N}\left(\frac{\pi\hbar c}{n_{\text{eff}}^{2}\lambda\epsilon_{0} V}\right)^{\frac{1}{2}}
\label{rabi_splitting}
\end{equation}
where $\bm{\muup}$ is the coupled dipole moment, $V$ is the cavity mode volume and $\hat{\textbf{E}}$ is a unit vector parallel to the polarization of the cavity electric field. The dipole moment of BChl $c$ is 5.48 D (Ref. \citenum{prokhorenko_exciton_dynamics_in}), and the cavity mode volume\cite{ujihara_spontaneous_emission_and} of the $q=2$ mode is calculated to be approximately 15 $\left(\frac{\lambda}{n}\right)^{3}$.

\end{document}